\documentclass[12pt,preprint]{aastex}

\newcommand{\kms}{km~s$^{-1}$}

\newcommand{\etal}{{\it et al.\/}}

\newcommand{\nsample}{24}

\begin{document}

\title{To Be or Not To Be: Very  Young Globular Clusters in M31
\altaffilmark{1}}

\author{Judith G. Cohen\altaffilmark{2}, Keith Matthews\altaffilmark{3}
\& P.~Brian Cameron\altaffilmark{2} }

\altaffiltext{1}{Based in part on observations obtained at the
W.M. Keck Observatory, which is operated as a scientific partnership 
among the California Institute of Technology, the University of 
California, and the National Aeronautics and Space Administration.}

\altaffiltext{2}{Palomar Observatory, MS 105-24,
California Institute of Technology, Pasadena, Ca., 91125, 
jlc(pbc)@astro.caltech.edu}

\altaffiltext{3}{California Institute of Technology, Downs Laboratory, 
MS 320-47, Pasadena, Ca., 91125, kym@caltech.edu}

\begin{abstract}

We present observations
made with the newly commissioned Keck laser-guide
star adaptive optics system (LGSAO) of 6 objects in M31 that are alleged 
in multiple recent studies to
be young globular clusters (GCs); all are supposed to have ages
$\leq$5  Gyr.
The resulting FWHM of the PSF core in our images is $\sim$70 mas.
The four youngest of these objects are asterisms; they are with
certainty not young GCs in M31.  Based
on their morphology, the two oldest  
are GCs in M31.  
While the M31 GCs with ages 5 -- 8 Gyr appear to be mostly genuine,
it appears that many
of the alleged very young GCs
in M31 are spurious identifications.   This problem will
be even more severe in studies of the GC
systems of more distant spiral galaxies now underway, for which
imaging at the spatial resolution of our
observations in M31 may not be adequate to detect sample
contamination by asterisms.

\end{abstract}


\section{Introduction}

M31 is the nearest large galaxy to our own, and as such plays a
a crucial role in many areas of extragalactic astronomy, both in
its own right and as a testbed for techniques for studying more
distant systems.  Early studies of its globular
cluster system (GCS) \citep[see, for example,][]{frogel},
suggested that the M31 GCS is very similar to our own, except
that it contains more members ($\sim$350 versus $\sim$150 members) and
extends to higher metallicity than does our own GCS, whose properties
are reviewed by \cite{harris96}. In this view, essentially 
all of the M31
globular clusters (GCs) are old, with typical ages of 10 -- 12 Gyr,
as is true of our own GCS, see, e.g \cite{rosenberg99}
and \cite{gcage}.

Recently, as more of the members of the M31 GCS have been studied
and additional spectra of these objects have been obtained,
there have been several papers asserting the existence of many
very young (age $\leq$1 Gyr), young (age 1 to 2 Gyr) and
intermediate age (3 to 6 Gyr) GCs in M31, in addition to a substantial
population of old GCs.  Among these studies, which discuss, to varying
degrees, the possibility of sample contamination,
are the work of
\cite{bar00}, \cite{bea05} (Bea05), \cite{puzia05} (P05),
\cite{bur05} (Bur05) and \cite{fusi05} (F05). 
Color-magnitude diagrams for a few of the brightest M31 GCs have been obtained
from HST imaging, see, e.g. \cite{rich} and \cite{williams}; the
latter focused on very young GCs, studying four of them in detail. 
Additional recent papers offer more sophisticated
interpretations of the presence of numerous very young and populous
clusters in M31, for example \cite{lee05}, using
these results to derive characteristics of the formation
of M31.

A young globular cluster is here defined as a
cluster of stars of essentially the same age and initial
chemical composition which has a total stellar mass similar
to that of the galactic GCs.  It must be centrally concentrated
and have a projected shape similar to that of galactic GCs.
It is an object which, with
the passage of time, when it reaches an age of 10 Gyr would
resemble a typical galactic GC.  Such populous young
clusters are not found in our galaxy, while
\cite{fusi05} claim that the youngest of 
these comprise 15\% of the whole of the M31 GCS.

In the present paper we show, from an admittedly small sample of
objects, that 2/3 of the sample we observed of purported very young
and young GCs in M31
turned out {\it{not}} to be GCs in M31.
It is thus likely that many of the
alleged youngest GCs in M31 are asterisms rather than globular
clusters.

\section{Observations}

We use the newly commissioned
laser-guide star assisted adaptive optics
\citep[LGSAO, ][]{wiz04} at the Keck II 10-m telescope on 
Mauna Kea, Hawaii.
The Near Infra-Red Camera 2 (NIRC2) 
(P.I. K.~Matthews) was used with a K' filter in the narrow field
mode.  Each pixel is then 0.010 arcsec on a side and the field
is approximately 10 x 10 arcsec.  This requires a natural
tip/tilt correction star which for best resulting image
quality must have R $< 16.5$ mag and be located less
than 45 arcsec from the object.  The LGS web page
(http://www2.keck.hawaii.edu/optics/lgsao/) 
provides details on the system performance, including
the image degradation when
these requirements are not met.  To be conservative,
we always obeyed these restrictions in designing our program.

LGSAO has the characteristic that the
resulting images have a Strehl ratio somewhat less than
that of the best achieved with natural guide star AO on
axis, but have  a much more uniform
PSF over the whole field than does natural guide star AO,
where the PSF degrades rapidly as one moves away from the correction
star. LGSAO permits the use of fainter guide stars
(required only for the tip/tilt correction) which can be
located somewhat further from the object of interest than
does natural guide star AO \citep{turb}.

We observed for four hours on each of two nights in August 2005. 
These nights
were marked by various technical problems, mostly at the beginning 
of each night, and by a shutdown of the laser for several hours
at the request of the U.~S. Space Command.  However, once the
laser was working, it ran without a major fault throughout
each night.  The natural
seeing ranged from 0.8 to 1.5 arcsec during the two nights. 
The NIRC2 exposures were
100 to 150 sec each, such long exposures in the thermal IR
being feasible with such a large telescope
only because of the very  small pixel scale
and resulting very low sky background signal.  
The longest total integration time was 1650 sec.
Individual
images show a  PSF even better than that HST,
with a core of between 60 and 80 mas.
The first diffraction ring is clearly visible on the best of these images.
There is also a diffuse halo component in the PSF which
for present purposes we ignore.

The LGSAO system has an acquisition camera with a field of 2
arcmin operating in the optical (effectively Johnson R). For each object, the 
pointing for the
tip/tilt star  was checked with this camera, then an offset
to the object of interest of less than
45 arcsec was made.  We used the first few integrations
on target
to identify and center up the object, then ran a script written 
by David Thompson
which made random moves within a specified box size to acquire
the subsequent images.  This moved the object around the field,
allowing a determination of the sky background from the
integrations on the object without separate sky exposures.  Dome flats
and bias frames were also acquired.
Since all our objects were within 3$^{\circ}$ of each other, the move
to and acquisition of subsequent targets was accomplished within
10 min.

\begin{figure}
\epsscale{.8}
\plotone{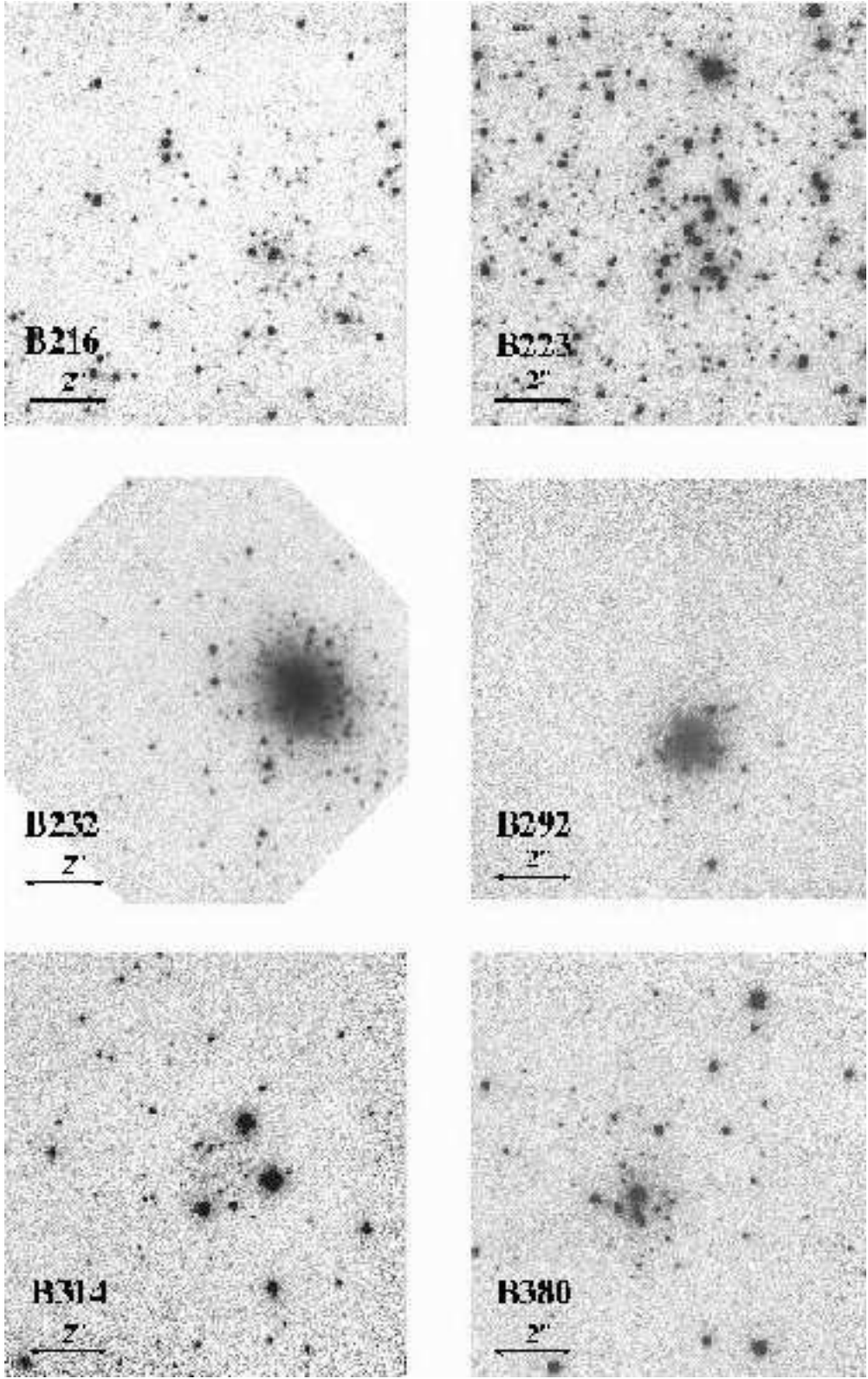}
\caption[]{\small LGSAO images from the Keck Telescope are shown for 6 putative
very young or young globular clusters in M31.  Based on their morphology,
only two of these are
actually GCs in M31.  These are K' images all oriented 
as indicated in the first
panel.  The field shown for each is approximately 10 arcsec on a side with a pixel
scale of 0.010 arcsec/pixel and FWHM with a core of 60 to 80 mas.
\label{fig_m31globs}}
\end{figure}

We followed standard reduction procedures for the set of images of each
object.
Each frame was flat-fielded, background subtracted and repaired for bad
pixels. We then produced a mask for each exposure to remove all detected
objects. We combined these masked images to produce an improved sky frame
for each of the 6 science targets for a second round of sky subtraction.
The optical distortion of the NIRC2
narrow field of view camera is relatively small.  It was corrected 
using algorithms derived from the pre-ship review documents
(available at http://alamoana.keck.hawaii.edu/inst/nirc2/) with the IDL
procedure provided by the Keck Observatory
(http://www2.keck.hawaii.edu/optics/lgsao/software/index.html).
We then shifted, co-added and trimmed the set of exposures
for each object, then divided by the number
of frames to produce the final images shown in
Figure~\ref{fig_m31globs}.

\section{Sample Selection}

We compiled a list of all  GCs in M31 with
ages less than 6 Gyr from the studies of Bea05, Bur05
and P05.  (The very recent extensive study of F05 was
not available when this was being done.)  We then
applied the restrictions for the LGS tip/tilt star described above.
We used the USNO-B1.0 Catalog \citep{monet03} to get a list of
potential tip/tilt stars with astrometric quality
positions.  Each such star (and each object) were then
checked on the Digitzed Sky Survey I frames to make sure that each
tip/tilt star was real, not a galaxy, and that it looked
stellar. 
We also eliminated objects with an obvious large spatially
varying background in the vicinity of the tip/tilt star (and the object)
due to concern over the stability of the LGSAO correction.

Object coordinates and photometry were taken from the recent update
of the Bologna catalog of M31 GCs of \cite{galet}.  Each of the GCs
observed was classified as a certain GC in M31 (classification flag 1) 
in this catalog
and each has a radial velocity consistent with
membership in M31.

Our initial list of candidates from these three recent papers
contained \nsample\ GCs.  After the requirements described above
 were imposed,
8 objects remained.  The 6 youngest of these were observed. 
The  ages of the two GCs in
the selected sample suitable for LGSAO imaging at the Keck Telescope
that were not observed  are 6.6 and 7.1 Gyr, older than any of those we
observed. 

The properties of our sample of M31 GCs observed
with LGSAO are given in Table~\ref{table_m31globs},
including their distances from the center of M31 along the major
(X) and minor (Y) axes of this galaxy.

\section{Results}

Figure~\ref{fig_m31globs} shows the resulting K' images taken
with LGSAO of the 6 very young or young GCs in M31.  It is immediately
apparent that only two of these are genuine GCs in M31.  These
correspond to the two objects in our sample with the oldest ages. 
Among the youngest objects in our admittedly small sample,  four
of the putative M31 GCs are clearly asterisms in M31; they definitely
are not genuine M31 GCs.  Two of these (B314 and B380) appear to be
very faint associations or clusters in M31 with superposed much brighter
objects.  These are  not genuine
M31 GCs; their integrated light as detected by normal seeing-limited ground
based observations would be dominated by the superposed stars rather
than by the very faint association/cluster and their spatial distributions are
not sufficiently centrally concentrated.
In all cases, the spurious objects as seen with a spatial
resolution typical of seeing-limited
ground based optical observations would appear
to be slightly extended objects.  It is not surprising that they
were picked up as GCs in M31.

The HST archive was searched for the \nsample\ objects in our original
sample prior to filtering for LGSAO.  WFPC2
images which included five of these were located.  One (B210,
not in our sample due to the lack of a suitable laser guide star) 
appears to be an asterism rather than a GC in M31.  The 
other four that were found there (B232, B311, B315, and B324)
all of which have an age $> 3$ Gyr, appear to be GCs in M31. 
One of these four (B232) is in our LGSAO sample; it has the appearance
of a GC in M31 there as well.  \cite{williams} presented HST images
for four ``young'' M31 GCs.  Ignoring duplications, they thus
verified that in addition B319, B342 and B368 are also genuine M31 GCs.
\cite{racine} (R91) imaged 109 halo candidate M31 GCs with the CFHT with 0.7 arcsec 
FWHM seeing to eliminate interlopers.

\section{Implications}

With the high spatial resolution of LGSAO at the Keck II Telescope
we have found that a large fraction of the putative very young and young
GCs we observed are not GCs in M31.  This suggests
that many
of the youngest GCs in M31 are asterisms rather than
M31 GCs.  Thus all the
theoretical papers explaining the surprising presence of numerous
very young GCs in M31 or using such to reach inferences
about the formation of M31 must be reconsidered, as must the
kinematic analyses based on the M31 GC radial velocities such
as \cite{morrison}.

The initial reconnaissance of the M31 GCS begun
more than 70 years ago by \cite{hubble}.  In recent years 
people have concentrated on detailed studies of larger samples of GCs in M31.
Recently Bea05, Puz05, Bur05, F05, among others, have
claimed the presence of a substantial number of
very young GCs in M31.  In photometric studies of the M31 GCS,
outliers which appear too red (i.e. older than 14 Gyr)
are rejected as distant background galaxies,
but outliers which appear very blue (i.e. very young) are usually not rejected.
Also the issue of
contamination of samples of GCs with asterisms
becomes more acute as one presses fainter.
While the claims of middle aged (5 -- 8 Gyr) GCs in M31
are not challenged by our observations, if there are no
or few very young M31 GCs, then at least some of the purported middle aged
GCs may easily be the outliers of the M31 GCS, once the sample
is restricted only to valid GCs in this galaxy.

\cite{cohen98} and \cite{cohen03} found that there
are no globular clusters with ages $\lesssim 8$ Gyr
in the Virgo gEs M87 and NGC~4472, a result extended by
\cite{puzia-vlt} to less luminous early-type galaxies with a somewhat
looser minimum age of 5 Gyr.  Elliptical galaxies have smooth
surface brightness profiles, at least outside their nuclei, and 
hence the study of their GC systems does not suffer from the
problem discussed here for the M31 GCS.

The potential confusion of GCs with asterisms is bound to become
more serious as the GC systems of more distant spiral galaxies
are explored.  Such projects are already underway by several 
groups at the largest ground-based telescopes.  Even the spatial resolution
offered by AO on a 10-m ground based telescope 
will at some point no longer be sufficient
to eliminate this possibility.  
Outliers may result from large observational errors caused by insufficient
integration time for such faint objects or
from sample contamination much larger than previously suspected. An example
of the former is the claim for young GCs in the Virgo elliptical NGC~4365
by \cite{larsen03}, which with better data, have subsequently been
shown to be old \citep{brodie05},
while the present paper illustrates the importance of the latter issue.
Caution is urged in interpreting the results of such studies.

\acknowledgements
We congratulate the Keck LGSAO team on the recent successful
commissioning of the LGSAO system.
The entire Keck user communities owes a huge debt to 
Jerry Nelson, Gerry Smith, and many other 
people who have worked to make the Keck Telescope
a reality and to operate and maintain the Keck Observatory. 
We are grateful to the W. M.  Keck Foundation for the vision to fund
the construction of the W. M. Keck Observatory. 
The authors wish to extend special thanks to those of Hawaiian ancestry
on whose sacred mountain we are privileged to be guests. 
Without their generous hospitality, none of the observations presented
herein would have been possible.
This research has made use of the USNOFS Image and Catalog Archive
operated by the United States Naval Observatory, Flagstaff Station
(http://www.nofs.navy.mil/data/fchpix/) and the Digitized
Sky Surveys produced at STScI based on photographic data obtained
using the Oschin Schmidt Telescope on Palomar Mountain and
the UK Schmidt Telescope.
We are grateful to the National Science Foundation for partial support 
under grant AST-0205951 to JGC.  

{}

\clearpage

\begin{deluxetable}{l crrrc l}
\tablenum{1}
\tablewidth{0pt}
\small
\tablecaption{Sample of M31 Globular Clusters Imaged With LGSAO
\label{table_m31globs}}
\tablehead{\colhead{ID} & \colhead{V,~~B--V\tablenotemark{a}} &
\colhead{$v_r$\tablenotemark{b}} & \colhead{Age} & 
\colhead{Refs.} & \colhead{X,Y\tablenotemark{c}}  & \colhead{Comments}  \\ 
\colhead{} & \colhead{(mag)} & \colhead{\kms} & \colhead{(Gyr)} &
\colhead{} & \colhead{(arcmin)} 
}
\startdata
B223 & 15.81,~0.15  & $-86$ & 0.5 & Bar00,J03,Bur05 & +26.4,$-3.8$ & asterism \\
B216 & 17.45,~0.20 & $-96$ &   0.5 & Bar00,J03,Bur05  & +26.+9,0.9  & asterism \\
B314 & 17.63,~0.69 &  $-318$ & $1.0\pm0.1$ &  Bar00,Bea04 & $-69.9,-10.8$ 
     & asterism\tablenotemark{d} \\
B380 & 17.12,~1.01  & $-121$  & $1.0\pm0.1$ & Bar01,Bea04,Bur05 &
   +58.5,$-2.1$  & asterism\tablenotemark{d} \\
B232 & 15.67,~0.72 & $-151$ & 2--5 & Bur05   & +12.5,$-17.5$ &   
    M31 GC, WFPC2\tablenotemark{e} \\
B292 & 17.00,~0.89 & $-331$ &  2.7--5.9 & R91,Bar01,Bea04,Bea05 & $-58.3,47.4$  &
    M31 GC \\
\enddata    
\tablenotetext{a}{from \cite{galet} }
\tablenotetext{b}{from \cite{huchra91}, \cite{galet} or \cite{bea04} }
\tablenotetext{c}{The coordinate system defined by \cite{huchra91} is used.}
\tablenotetext{d}{Very faint loose association/cluster with superposed much
brighter stars.}
\tablenotetext{e}{HST image 
   taken 9/1998 as part of an archive program for parallel WFPC2 images  
   in the neighborhood of bright galaxies, PI S. Casertano}
\end{deluxetable}

\clearpage

\end{document}